\documentclass{article}

\usepackage[preprint]{neurips_2024}

\usepackage[utf8]{inputenc} 
\usepackage[T1]{fontenc}    
\usepackage{hyperref}       
\usepackage{url}            
\usepackage{booktabs}       
\usepackage{amsfonts}       
\usepackage{nicefrac}       
\usepackage{microtype}      
\usepackage{xcolor}         
\usepackage{graphicx}

\title{Improving text-conditioned latent diffusion for cancer pathology}

\author{%
  Aakash M. Rao \\
  Department of Computer Science\\
  Ashoka University\\
  Sonipat, India, 131029 \\
  \texttt{mrao.aakash@gmail.com} \\
  \And
  Debayan Gupta \\
  Department of Computer Science\\
  Ashoka University\\
  Sonipat, India, 131029 \\
  \texttt{debayan.gupta@ashoka.edu.in} \\
}

\begin{document}

\maketitle

\begin{abstract}
  The development of generative models in the past decade has allowed for hyperrealistic data synthesis. While potentially beneficial, this synthetic data generation process has been relatively underexplored in cancer histopathology. One algorithm for synthesising a realistic image is diffusion; it iteratively converts an image to noise and learns the recovery process from this noise \citep{art_57}. While effective, it is highly computationally expensive for high-resolution images, rendering it infeasible for histopathology. The development of Variational Autoencoders (VAEs) has allowed us to learn the representation of complex high-resolution images in a latent space. A vital by-product of this is the ability to compress high-resolution images to space and recover them lossless. The marriage of diffusion and VAEs allows us to carry out diffusion in the latent space of an autoencoder, enabling us to leverage the realistic generative capabilities of diffusion while maintaining reasonable computational requirements. \cite{art_43} and \cite{art_38} build foundational models for this task, paving the way to generate realistic histopathology images. In this paper, we discuss the pitfalls of current methods, namely \citep{art_38} and resolve critical errors while proposing improvements along the way. Our methods achieve an FID score of $21.11$, beating its SOTA counterparts in \citep{art_38} by $1.2$ FID, while presenting a train-time GPU memory usagereduction of $7\%$.
\end{abstract}

\section{Introduction}
Various machine learning and computer vision algorithms rely heavily on the largeness of datasets and the representation of classes to learn generalizability. This process is complicated in a field such as pathology for various reasons, one of which is the severe scarcity of high-quality labelled data \citep{art_59}. This limitation directly impacts the ability of models to learn different classes efficiently, especially in highly heterogeneous data such as pathological data. One manner of dealing with class imbalance has been in the form of augmentations to the original input data in the form of rotations, flips, translation, scaling, brightness and contrast adjustments, noise injection, blur, colour jitter, etc. \citep{art_60,art_61}. Some of these transformations are safe to execute and will not alter the image in a manner that changes its structural features. However, some of these transformations can alter the structural features of a given image. For example, noise injection and brightness and contrast adjustments in pathology images can induce artefacts that could have unintended consequences \citep{art_62,art_63}. Building models that can generate synthetic pathological images effectively has the potential to circumvent these transformations. It is important to note that we are considering an ideal model in this case and illustrating the importance of synthetic data. Further, synthesising further data could help expand our understanding of rare cancer types with limited sample availability.\\

Further, increased data also allows for better-automated image analysis frameworks. While the computational advantages of synthetic images are vast, a lesser-discussed advantage of artificial images, especially in a field such as pathology, is in the education sector. Generating images that are true to their label (structurally concordant to the label that defines them) can be used to train pathologists around the world with a much more extensive and diverse set of samples \citep{art_64,art_65}. This could be especially useful in training junior pathologists and trainees in identifying rare and often misdiagnosed cancer types \citep{art_64,art_65}. The authors of \citep{art_38} suggest a unique approach. Inspired by the work done by \citep{art_39}, PathLDM combines the representation capabilities of VAEs with the strong, realistic generative capabilities of DMs to produce realistic, high-resolution pathology images \citep{art_38}. This work, while state-of-the-art, has certain elements that could hinder performance. In this work, we address these pitfalls, primarily the summary generation process, the data structuring, and the work’s overall reproducibility. In doing so, we provide performance and memory usage improvements.

\section{Related Works}
There exists comparatively limited literature exploring the application of generative models for synthetic image generation, specifically for cancer histopathology. \citep{art_32} propose a novel unified method to train and refine a GAN-based generation pipeline along with a task-specific CNN, boosting its performance using on-the-fly generated adversarial samples. The paper outlines a generation technique that produces images with desired characteristics “such as the locations and sizes of the nuclei, cellularity, and nuclear pleomorphism” \citep{art_32}. They use authentic images as a reference to generate synthetic images in the reference style using GANs, which is similar to other style transfer techniques used in different domains \citep{art_32}. This approach has been similarly explored by \citep{art_34}, who present GAN-based image generation as an alternative to standard data augmentation processes.\\

Subsequently, the authors of \citep{art_33} design an attribute-guided GAN for histopathology image synthesis. Here, they can demonstrate that their work’s unique ability to use multi-attribute annotation to control image synthesis significantly improves the image synthesis process \citep{art_33}. Further, they develop a unique Skip Layer Channel-wise excitation and a reconstruction loss in the discriminator to better capture global context along with this mentioned multi-attribute annotation \citep{art_33}. A more straightforward approach was adopted by \citep{art_35} in their work to produce class-conditioned histopathology images. The authors of \citep{art_36} are among the initial few to demonstrate the benefits of applying diffusion to the space of histopathology images. They develop a nuclei-aware diffusion approach that produces synthetic images conditioned upon semantic instance maps consisting of up to six different types of nuclei \citep{art_36}. \\

The application of diffusion models in pathology is further limited. \citep{art_52} present the first application of DMs to pathology, working at a pixel level to generate class-conditioned diffusion. The authors introduce a pipeline for LDMs to generate synthetic histopathological images for Low-Grade Gliomas (LGGs). They directly use patches of WSIs as inputs to the model along with a conditioning vector (t) that defines the class of the input patch. This work was limited due to the high dimensionality of input images fed into the DM, requiring significant computation. Similarly, \citep{art_53} evaluated the performance of DMs on a different class-conditioned problem and illustrated a considerable performance improvement over conventional GAN-based image synthesis methods. While the process of \citep{art_53} was built upon the foundation of \citep{art_52}, the latter proposes a vital improvement. The authors of \citep{art_53} proposed working with Latent Diffusion Models (LDMs) as an alternative to diffusing directly in the image space. This presented significant improvements over the previous work but was still limited to class-conditioning labels. Additionally, both \citep{art_52} and \citep{art_53} work with relatively small datasets, which may be confounding factors in successfully learning the image synthesis task.

\section{Methodology}
\subsection{PathLDM}

\begin{figure}
  \centering
  \includegraphics[width=\textwidth]{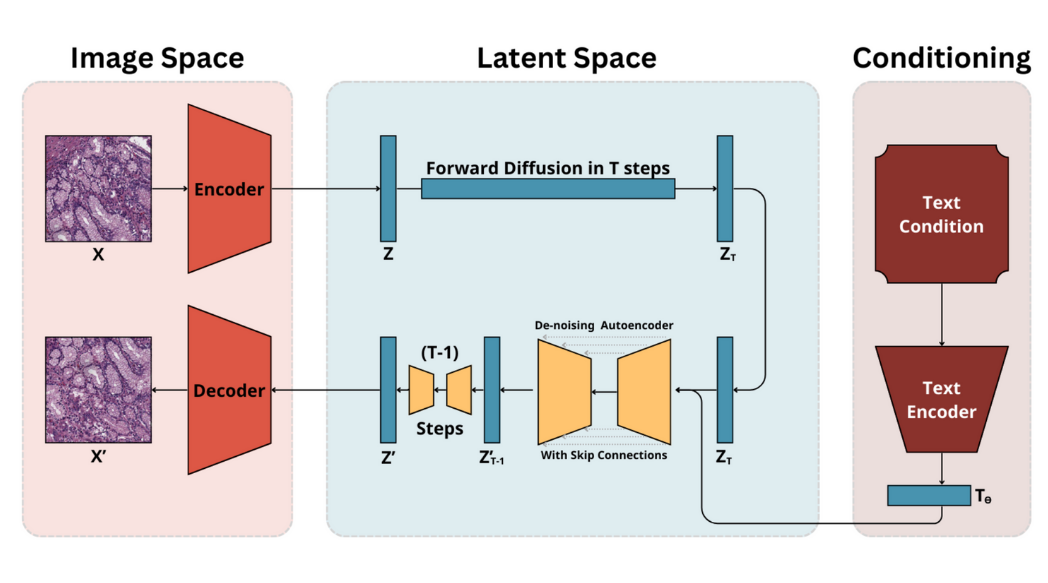}
  \caption{Network diagram showing diffusion acting in the latent space produced by a variational autoencoder. Also highlights the text embedding framework working before the first step of reverse diffusion}
\end{figure}

The architecture of PathLDM is inspired by the work done in \citep{art_44}, where the ability of a VAE to encode images and recover them from a latent space was leveraged to allow for high-dimensional image synthesis. The VAE’s unique ability to compress a given input to a tighter space generates a low-dimensional representation of the originally high-dimensional image. This enables us to successfully carry out and learn a forward and reverse diffusion process within this latent space, allowing for a computationally feasible high-dimensional image synthesis process. The VAE used here was borrowed from the work \citep{art_43}, pre-trained on the ImageNet dataset \citep{art_45}. This VAE, based on the work by \citep{art_45}, consists of a combination of perceptual loss and a patch-based adversarial objective borrowed from the GAN architecture \citep{art_46,art_47}. These improvements allow for better local realism and avoid the issues with blurred data generations when using L1 and L2 objectives. As we saw earlier, the encoded space is a factor of f smaller than the original image space, where f is known as the downsampling factor. The authors of \citep{art_38} empirically found that f = 4 is the most suitable value that balances the tradeoff between loss of perceptual information and increasing computational complexity \citep{art_43}. \\

The DM implemented uses the latent space representation of the original image to iteratively noise the image and learns the reverse denoising process \citep{art_43}. The reverse process consisted of iterative use of a time-conditioned u-net architecture, first developed by \citep{art_48} for image segmentation. The text condition served as an additional input to the first layer of the reverse diffusion process using a text encoding mechanism \citep{art_43}.

\subsection{Data }
The data for PathLDM is wholly sourced from The Cancer Genome Atlas (TCGA) and specifically the TCGA Breast Invasive Carcinoma Collection (TCGA-BRCA). This cohort consists of data belonging to approximately 1098 cases (Patients) across different parts of the continental United States. The data used for this work consists of the pathological whole slide images (WSIs) and the pathological report corresponding to the given case. The reports used as the basis for this work were provided by \citep{art_40}, who carried out the digitisation process for over 10,000 reports present on the TCGA GDC. This work is intended to better benchmark LLMs attuned to pathological reports; however, we, and the authors of \citep{art_38}, utilise these reports as a source of information to create text conditions for our pathology images \citep{art_38}\citep{art_40}. The text condition consisted of a summary of the pathology report, with tumour and TIL scores. The summary of the pathology report was generated using GPT3.0, and the Tumour and TIL scores were generated by \citep{art_42} and \citep{art_41}, respectively. The summaries were embedded using Contrastive Language-Image Pre-Training (CLIP). Since CLIP’s embedding frame is a maximum of 77 tokens, the halves of the text condition were individually embedded and subsequently concatenated. 

\subsection{Pitfalls} 
The work presented in \citep{art_38} is robust and a foundational model for applying diffusion in high-dimensional image synthesis. That said, it does have certain pitfalls and limitations. The work’s first and most crucial issue is its need for immediate reproducibility. The code given in the repository of \citep{art_38} is vast and heavily borrowed from the authors of  \citep{art_43}. However, there exists a few issues:
\begin{itemize}
    \item The environment supplied with the work is dated, with the use of deprecated packages limiting the recreation of this environment.
    \item The code borrowed from \citep{art_43} implements the powerful use of distributed computing to allow for training across multiple GPUs using a gradient accumulation technique. However, in doing so, it fails to account for training scenarios involving single-GPU systems.
    \item This, coupled with a seemingly arbitrary use of mixed precision and full precision computation, leads to immediate errors preventing reproducibility.
\end{itemize}

The text prompts presented to the LDM model as a condition of image generation are essential. It is necessary to ensure that they are indicative of image content in the best manner possible. As discussed earlier, information from the patient’s case is a large part of the summary; around 150 of the 154 possible tokens are dedicated to this information. While patient information, such as the cancer subtype, is essential, it may not necessarily be relevant to synthetic image generation. For example, we sample two patches from independent WSIs; both consist primarily of fat. Additionally, running the tumour and TIL prediction pipelines yielded matching scores of “Low tumour;” and  “Low TIL;” However, our summaries may be different, where one corresponds to invasive ductal carcinoma (IDC), while the other corresponds to invasive lobular carcinoma (ILC). This leads to two fundamental difficulties:
\begin{itemize}
    \item  The caption is weighted to a summary and does not indicate the image fed into the algorithm.
    \item Multiple similar images, i.e., of fat, have a large body of labelling that is very heterogeneous.
\end{itemize}
This may lead to inferior performance in the model due to the simple reason of labelling inconsistencies. \\

Further, while the original prompts have been provided, no code exists on the repository that allows users to experiment with customisable prompts and further tweak this pipeline’s settings in any way. While the train-test split of the original data has been provided, there needs to be insight into the method used to generate them. This, again, limits the results presented in the original paper, as further experimentations with random splits are required to confirm the results mentioned. Further, the initial token size of 150 tokens used to generate the summaries is arbitrary. No rationale has been provided as to why this token length was considered. Suppose one includes the four additional tokens related to tumour and TIL scores. In that case, the total token length comes to 154 tokens, which allows for a maximum of 2 units in the positional embeddings. We assume that this is the reason for the parameter chosen.\\

In addition to the need for more reproducibility of the summarisation workflow, the lack of experimentation with captions of various lengths is another limitation. Examining any one of the summarised captions allows us to see that the generated 154 token captions contain a combination of information that could be potentially relevant to the image patch being learnt. However, in some situations, we also see that information, such as the filler text, may need to be more relevant to the image generation process. Furthermore, a component of the information is of marginal relevance to the reconstructed image and falls in the grey area of whether it should be included. Examining the captions used as text-conditions could be a key area of improvement. Therefore, revisiting the summary generation process may improve the model’s outcome, i.e., an enhanced FID score and agreement between the image patch and text condition.\\

\subsection{Improvements Presented}

\begin{figure}
  \centering
  \includegraphics[width=\textwidth]{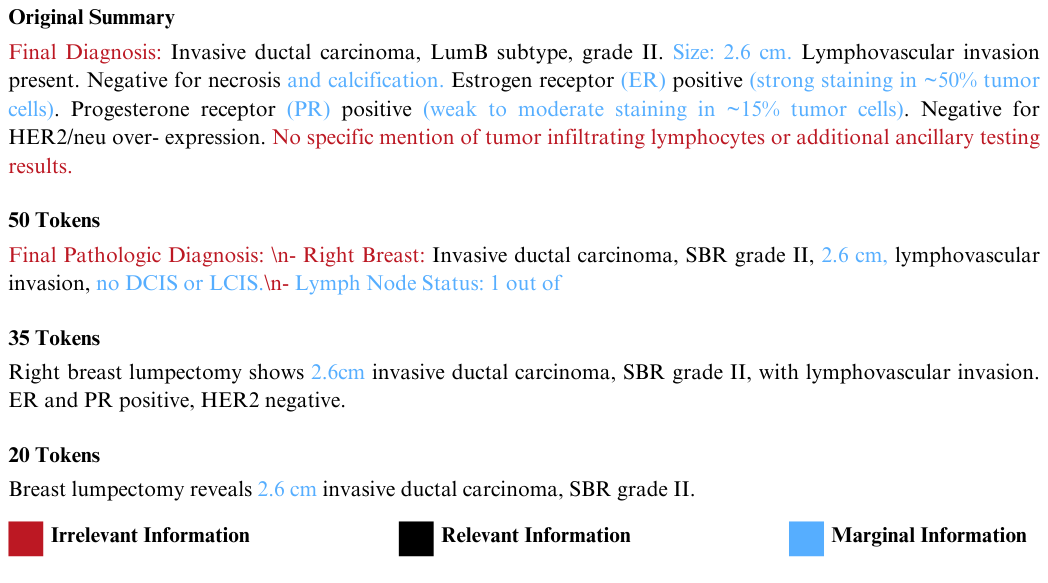}
  \caption{Three token lengths with varying information (top to bottom) 154-token summary with a mix of irrelevant and marginal information, and reduction of irrelevant and marginal information between 50 and 35 tokens, further reduction of relevant information in 20-token summary}
\end{figure}
The first task of any open-source work is to ensure reproducibility. As seen in the pitfalls section, we first had to deal with a lack of functionality in the work presented by \citep{3}. Through our experience implementing other unrelated works, we decided first to understand if the errors resulted from the code itself or the environment we had set up. After carefully exploring the log files, we were able to isolate the original set of errors to the version of PyTorch-Lightning, specifically to the incorrect channel declaration. This issue was resolved by declaring the required information in the environment YAML files. \\

As mentioned earlier, \citep{art_38} had heavily borrowed code from \citep{art_43}, built to work on a distributed GPU system. \citep{art_38} also implemented this similarly, leading to a need for experimentation on a single GPU machine. This, coupled with older versions of specific packages and the inconsistent use of mixed and full-precision, caused further errors. These errors were arbitrary and were reflected in the logs as issues with PyTorch-Lightning, causing significant uncertainty. This issue was resolved in a thorough examination of the work by \citep{art_38} and \citep{art_43}, along with insight and conversations from Srikar Yellapragada, the author of \citep{art_38}. We found that some alternative cases needed to be appropriately addressed when a multi-GPU setup failed due to a lack of multiple GPUs. This resulted in full-precision use where mixed-precision was required. These alternative cases were triggered when we reimplemented the work by \citep{art_38} on our single GPU machines. The lack of appropriate error flags further complicated this issue. We would finally downcast elements at these stages and reproduce the work. \\

The lack of a summarisation pipeline presented a unique challenge in addressing this issue. It allowed for more customisation regarding summary length and other alternative prompts that future researchers would like to explore. To do this, we developed a workflow integrated with the OpenAI API that allows for the token-length parameterised generation of summaries. We have thoroughly tested this pipeline with different parameters and prompts. Considering the dependency on an API that may not always return appropriately for every request, we also integrated a re-generation functionality. This gives users a more functional approach to addressing relevant corner cases. We tested this pipeline with the same prompt sequence in \citep{art_38}. This integration of the API also allowed us to allow compatibility with other, newer GPT models, such as GPT4. However, we could not test the workflow using different models due to financial constraints, so we used gpt3.5-turbo, as mentioned by the authors in \citep{art_38}. 
As discussed earlier, upon further examination of the generated summaries and developing the summarisation workflow, we found merit in understanding a better way to create summaries. We aim to maximise the information relevant to the presented image and minimise irrelevant information that could act as potential noise. To do this, we experimented with two areas. First, we changed the final prompt present in the original sequence to ask for more targeted information in the summary. Secondly, we experimented with generating summaries with different token lengths to view the difference between the information presented in them. The results of this are presented in the next section. We evaluated summaries at various lengths and found that 20, 35, and 50 token summaries encapsulated all relevant information. The summary of 20 tokens was too short, leading to pertinent information withheld, while the 50-token summary was too large, with irrelevant information introduced. The 35-token summary was ideal as it balanced the amount of information and length of the summary. \\

We additionally provide a two-pronged workflow to generate synthetic images and evaluate them. In the first part, we used the starting point provided by \citep{art_38} and \citep{art_43} and utilised DDIM sampling to efficiently and speedily generate samples. DDIM sampling allows us to quickly and efficiently generate numerous samples, allowing for en masse text-conditioned generation. Based on the information in \citep{art_38}, we used 50 steps and a scale of 1.75, which was found to minimise the FID score. We used these settings and this workflow to generate images for all the models we trained, using the test set of just over 2 lakh captions. We also highlighted the requirement of a formal evaluation procedure for the models with a reproducible framework. To do so, along with the prediction and data generation workflow discussed previously, we added support for FID score calculation using the PyTorch-FID package from GitHub, available on PyPI \citep{art_52}. \\

Addressing the issues we presented in the previous section is a complex process. We could not effectively address this due to the time and logistical constraints. We would have liked to look at various approaches that could combine the two halves of the embeddings or even use alternative embedding strategies that avoid splitting the original information. Nonetheless, we did implicitly experiment with this, as our variation of token length allows us to compare the effectiveness of conventional embeddings to cyclic positional embeddings mentioned in \citep{art_38}. All the models were trained on a single NVIDIA A5000 GPU for 150 GPU hours, amounting to approximately 460000 iterations, with a batch size of 32. The models were evaluated and compared in terms of FID scores.  We hypothesised that the different lengths of text conditioning would have varied impacts on the generative performance of the model. Further, specifically, we hypothesised that the 35 token summaries would be the most appropriate text condition as it would lead to maximal information relevant to the patches being generated, while the 20 token summaries would be the worst. The 50 and 154 token summaries would lead to similar performance. We also posit a possible reduction in train-time GPU memory usage.\\

\section{Results}
\subsection{Pitfalls} 
\begin{figure}
  \centering
  \includegraphics[width=\textwidth]{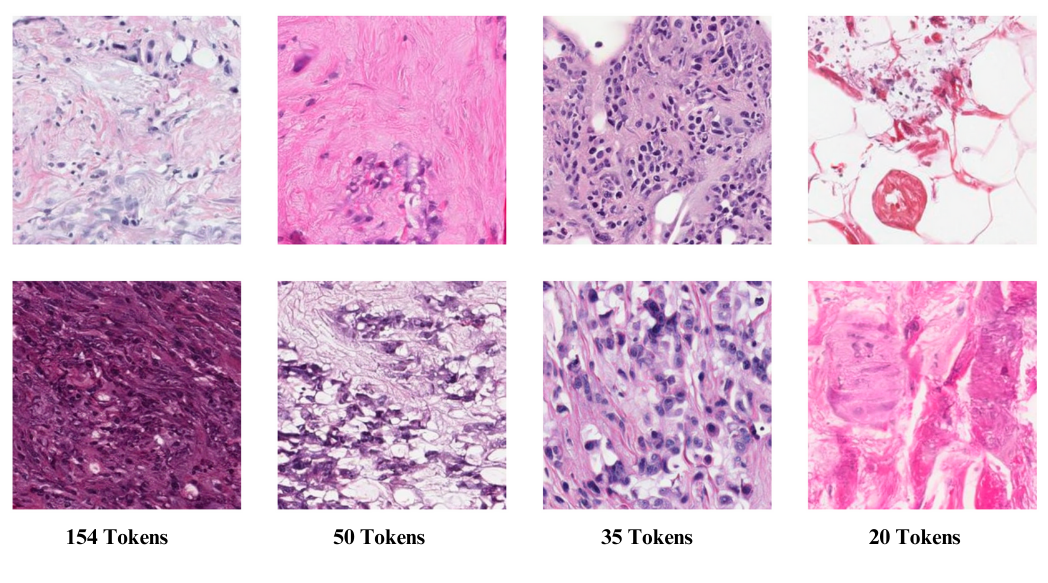}
  \caption{Synthetic images generated from randomly sampled summaries from the test set (left to right) in reducing summary length}\label{fig3}
\end{figure}
Out of the four token lengths, the model trained on the 20-token summaries was the worst performer, with an FID score of $24.01$. The best performer was the 35-token summaries, with an FID score of $21.11$. The 50 token summaries had the second-best performance, with an FID of $21.51$. The original summaries of 154 tokens were the second-worst performers, with an FID of $22.39$. The original model utilised $14.63$ GB of train-time GPU memory, while our best-performing model utilised $13.6$ GB. The 20 and 50 token models utilised $13.48$ GB and $14.07$ GB, respectively. Our methods outperform prior state-of-the-art models for text-conditioned pathology synthesis such as Medfusion \citep{art_53}, Stable Diffusion \citep{art_43}, and \cite{art_52}. Therefore, our best-performing model presents a $1.2$ FID improvement over the current state-of-the-art while presenting $7\%$ lesser GPU memory usage during training. Further 

\begin{table}
  \caption{Results of our 35-token model compared to other SOTA models}
  \label{sample-table}
  \centering
  \begin{tabular}{lll}
    \toprule
    \textbf{Method} & \textbf{FID Score}  \\ \midrule
    \citep{art_52}   &  105.81\\
    Medfusion \citep{art_53}        &  39.49\\
    Stable Diffusion \citep{art_43} &  30.56\\
    PathLDM \citep{art_38}          &  22.39\\
    Our Best                        &  \textbf{21.11}\\
    \bottomrule
  \end{tabular}
\end{table}

\begin{table}
  \caption{Summary of various token lengths and their corresponding FID and train-time GPU memory usage}
  \label{sample-table}
  \centering
  \begin{tabular}{lll}
    \toprule
    \textbf{Token Length} & \textbf{FID Score} & \textbf{GPU memory Usage (GB)}  \\ \midrule
    20  & 24.01 & \textbf{13.48} \\
    35  & \textbf{21.11} & 13.6  \\
    50  & 21.51 & 14.07 \\
    154 & 22.39 & 14.67 \\
    
    \bottomrule
  \end{tabular}
\end{table}

\section{Discussion}
In this work, we have shown the effect of various summary lengths on the image generation performance of latent diffusion models. We show that optimising the summary generation process to produce succinct summaries that maximise patch-related information is vital to generating more realistic images. Our 20-token models are the worst performers, as we have demonstrated a lack of relevant information. Similarly, the original models by \citep{art_38} also perform poorly owing to the induction of irrelevant details that could act as noise. The 35-token and 50-token summaries maximise information of patch relevance, leading to their outperforming SOTA. In the original paper, \citep{art_38} reports an FID of nearly 7.34. However, the lack of reproducibility and other issues encountered during the work led to one questioning their validity. Therefore, for comparison of performance, we only consider the variant of PathLDM reproduced by us, along with the other methods such as Medfusion \citep{art_53}, \citep{art_52}, and Stable Diffusion \citep{art_43}. Due to a lack of pathological context, further work would be necessary in the embedding domain to understand if CLIP generates the most appropriate embedding. While the authors of \citep{art_38} have experimented with PLIP as an alternative to CLIP, there is merit in testing alternative embedding strategies with other interpretations of the information in the pathological reports. Lastly, there is merit in including more patch-level information, beyond TIL and tumour, such as fat and stromal distribution, that could help with better realism.




\begin{ack}
This research has been supported by the Mphasis M1 foundation and the Ashoka High-Performance-Computing (HPC) team. We also thank Dr Rintu Kutum for the computational and infrastructure support. The results shown here are in whole or part based upon data generated by the TCGA Research Network: \href{https://www.cancer.gov/tcga.}{https://www.cancer.gov/tcga.}
\end{ack}

\subsection*{Data and Code}
All the code and data for this work is available on the project repository at \href{https://github.com/mraoaakash/mraoaakash-capstone-thesis}{GitHub}

\bibliographystyle{plainnat}
\bibliography{main}


\appendix



\end{document}